\ifx \doiurl    \undefined \def \doiurl#1{\href{http://dx.doi.org/#1}{\textsf{DOI}}}\fi
\ifx \adsurl    \undefined \def \adsurl#1{\href{http://adsabs.harvard.edu/abs/#1}{\textsf{ADS}}}\fi
\ifx \arxivurl  \undefined \def \arxivurl#1{\href{http://arxiv.org/abs/#1}{\textsf{arXiv}}}\fi

\documentclass[pdftex,12pt]{article}
\usepackage{natbib}
\usepackage[T1]{fontenc}
\usepackage[english]{babel}
\usepackage[breaklinks,colorlinks]{hyperref}
\hypersetup{
citecolor=blue,
urlcolor=blue
}
\usepackage{graphicx}        
\usepackage{color}           
\def\arcsec{$^{\prime\prime}$}
\usepackage{amssymb}

\usepackage{textcomp}
\usepackage{gensymb}
\usepackage[font={small}]{caption}

\title{\vskip-2cm
Oscillations Above Sunspots and Faculae: Height Stratification and Relation to Coronal Fan Structure}

\author{N.I.~Kobanov, D.Y.~Kolobov, and A.A.~Chelpanov\\
    \small{Institute of Solar-Terrestrial Physics} \\
    \small {of Siberian Branch of Russian Academy of Sciences, Irkutsk, Russia} \\
    \small {email: \url{kobanov@iszf.irk.ru}}
    }
\date{\small{[
{This article was firstly published in {\textit{Solar Physics}} \href{https://dx.doi.org/10.1007/s11207-014-0623-6}{DOI}}]}}

\begin{document}
\maketitle
\begin{abstract}
Oscillation properties in two sunspots and two facular regions are studied using
Solar Dynamics Observatory (SDO) data and ground-based observations in the
Si\,\textsc{i}\,10827\AA\ and He\,\textsc{i}\,10830\AA\ lines.
The aim is to study different-frequency spatial distribution characteristics above
sunspots and faculae and their dependence on magnetic-field features and to detect
the oscillations that reach the corona from the deep photosphere most
effectively.
We used Fast-Fourier-Transform and frequency filtration of the
intensity and Doppler-velocity variations with Morlet wavelet to trace the wave
propagating from the photosphere to the chromosphere and corona.
Spatial distribution of low-frequency (1\,--\,2 mHz) oscillations outlines well the fan-loop
structures in the corona (the Fe\,\textsc{ix} 171\,\AA\ line) above sunspots and faculae. High-frequency
oscillations (5\,--\,7 mHz) are concentrated in fragments inside the photospheric
umbra boundaries and close to facular-region centers.
This implies that the upper parts of most coronal
loops, which transfer low-frequency oscillations from the photosphere, sit in the
Fe\,\textsc{ix} 171\,\AA\ line-formation layer.
We used dominant frequency \textit{vs.}
distance from barycenter relations to estimate magnetic-tube inclination angle
in the higher layers, which poses difficulties for direct magnetic-field
measurements. According to our calculations, this angle is ${\approx}$40$\degree$ in the 
transition
region around umbra borders. Phase velocities measured in the coronal loops'
upper parts in the Fe\,\textsc{ix} 171\,\AA\ line-formation layer reach 100\,--\,150\,km\,s$^{-1}$ for
sunspots and 50\,--\,100\,km\,s$^{-1}$ for faculae.

\end{abstract}

\section{Introduction}
     \label{S-Introduction}
\par The intensive study of oscillations observed in the solar
atmosphere has lasted half a century. Many new facts about the origin and
properties of oscillations have been obtained. One of the relevant problems
is the interaction mechanism between the solar magnetic field and
oscillations. According to early studies, the oscillation properties
in the magnetic-field regions differ significantly
from those observed in non-magnetic ones \citep{how1968sp,balthasar1984,lites1984a,kob1985}.
\par The most noticeable magnetic-field regions on the Sun are sunspots
and faculae. Magnetic structure of a single regularly formed sunspot 
has circular symmetry. Magnetic field is vertical in the center of a
sunspot. The inclination increases with distance from the barycenter, and the field becomes
almost horizontal in the outer penumbra. This feature makes sunspots
the most attractive objects to study the relationship between the oscillatory-wave
process properties and the magnetic field topology.
\par The study of oscillations in the lower layers of the solar atmosphere
has revealed the decrease in dominant frequency with
distance from the sunspot center \citep{rimmele1995,sigwarth1997,kob2000sp,kobmak2004}.
A peculiar property of sunspots is the existence of running penumbral waves (RPW),
which are readily detected in the sunspot chromosphere
\citep{beckers1969uf,giovanelli1972,zirin1972}. Two scenarios were
proposed to interpret RPWs. According to the first one -- the trans-sunspot
wave scenario -- the waves propagate from sunspot umbrae to penumbrae
along an almost horizontal trajectory
\citep{alissandrakis1992,tsiropoula2000,tziotziou2004b,tziotziou2006a}.
According to the second one -- the visual pattern scenario -- the waves
propagate from the lower layers to the upper ones along a path
with different angles. When observing the chromosphere, one
sees an illusion of horizontal propagation
\citep{rouppe2003,bogdan2006a,bloomfield2007a,kobkol2009}.
\par Oscillations in sunspots and faculae have combined and individual
properties. For example, studies of the line-of-sight (LOS) velocity
showed that five-minute band oscillations dominate in the photosphere of
sunspot umbrae \citep{how1968sp,bhatnagar71b,thomas1982Natur,balthasar1984}
and faculae \citep{orrall65,Howard67,bhatnagar71}; the amplitude
of the oscillations is suppressed relative to the undisturbed
regions. Three-minute-band oscillations dominate in the chromosphere above
spot umbrae \citep{beckers1969uf,kneer1981,lites1984a,lites1986a,ZhugSych},
while the facular chromosphere preserves five-minute oscillations and even reveals
lower-frequency oscillations
\citep{orrall65, Howard67,teske74,blo1971,woods80,balthasar90,kobpul2007}.
\par The present opinion is that coronal five-minute oscillations are observed
mainly above faculae and the chromospheric network
\citep{dem2000aa,cen2006ASPC,vec2007aa},
while three-minute oscillations in the solar corona are related to sunspots
\citep{oshea2002a,doy2003sp}. Three-minute oscillations above sunspots in the
microwave range were observed with a 50-second delay compared with the
chromospheric oscillations\citep{AbMax2011}.
\par Now excellent Solar Dynamics Observatory (SDO) data allow us to study mode properties
in sunspots and faculae at different heights from the
photosphere to the corona \citep{RezSh2012,rez2012ApJ,kob2014ARep}.
\par Altitudinal cuts of the frequency spatial localization
allow us to trace the path of wave perturbations. In the future,
this technique can be used to determine magnetic-field inclination in the
transition zone and the low corona using an alternative method \citep{jess2013}.
\par Fan structures are often observed in the corona above sunspot and faculae.
These structures are seen especially clearly in the Fe\,\textsc{ix} 171\AA\ line. It is possible
to determine the frequency of the waves traveling along fan
structures by comparing the spatial distribution (power localization)
of selected frequency waves with the Fe\,\textsc{ix} 171\AA\ line images of these regions.
The relation between fan structures and waves is
the subject of many studies. \cite{marsh2006} found that three-minute
umbral oscillations propagate directly into coronal loops.
Similar conclusions have been made by \cite{jess2012apj}, who
demonstrated that coronal loops are anchored in the photospheric
umbral dots with enhanced intensity of three-minute oscillations.
Earlier \cite{bry2004sp} showed that three-minute oscillations in the corona
were confined to the narrow domain that corresponded to the sunspot
umbra boundary at the photospheric level. These oscillations are
suppressed in fan structures. \cite{wang2009apj} revealed
12- and 25-minute oscillations in coronal fans above sunspots
and identified them as slow magnetoacoustic waves. \cite{kob2013aa}
supported the conclusions made by \cite{bry2004sp} concerning three-minute
oscillations, and detected waves with periods of 12\,--\,15 minutes
propagating from the penumbral photosphere to the coronal fans.

\section{Methods}
\par We used both space- and ground-based telescope observations that cover the
same active regions at the same temporal intervals.
SDO has three instruments onboard: the Atmospheric Imaging Assembly (AIA), 
the Helioseismic and Magnetic Imager (HMI)
and Extreme Ultraviolet Variability Experiment (EVE),
more details can be found in \cite{lem2012sp,sch2012sp,woods2012sp}. The first
provides data in a wide range of UV spectral lines, which cover heights
from the photosphere to the corona. The intensity image series have cadences of 12 seconds
for all but two lines: the 1600\,\AA\ and 1700\,\AA\ bands' cadence is 24 seconds. We
chose four bands for the analysis. The coronal heights are represented by the
Fe\,\textsc{ix} 171\,\AA\ and Fe\,\textsc{xii}, \textsc{xxiv} 193\,\AA\ lines. The
other two are the continuum 1700\,\AA\ (the upper photosphere) and
He\,\textsc{ii} 304\,\AA\ (the transition region) lines, whose formation heights are
the closest to those of the Si\,\textsc{i} 10827\,\AA\ and He\,\textsc{i} 10830\,\AA\
lines, which we used in the ground-based observations.

\par The second instrument, HMI, provides intensity, Doppler velocity, and
magnetic field data obtained using the Fe\,\textsc{i} 6173\,\AA\ line with a 45 second
cadence. This line is formed in the photosphere at a height of 100\,--\,150\,km
\citep{fleck2011sp,parnell1969sp}.

\par The ground-based observations were obtained with the solar telescope at the Sayan
Solar Observatory located at an altitude of 2000\,m. The telescope is elevated six meters
above the ground and is equipped with a special system to suppress atmospheric turbulence
\citep{ham1973}. The telescope contains a coelostat, with the effective diameter of
its main mirrors being 800\,mm and a guiding system that keeps the Sun's image on the
spectrograph slit with an accuracy of one arcsec. We used a Princeton Instruments
RTE/CCD camera (256${\times}$1024). One element corresponded to a spatial resolution of
0.23 arcsecs along the entrance slit of the spectrograph and to 7\,m\AA\ in the
direction of the spectrograph dispersion. The observations provided a series of
spectrograms with a cadence of 3 seconds containing the He\,\textsc{i} 10830\,\AA\ and
Si\,\textsc{i} 10827\,\AA\ spectral lines. For details, see
\cite{kob2013sp}.

\begin{figure}
\centerline{
\includegraphics[width=7cm]{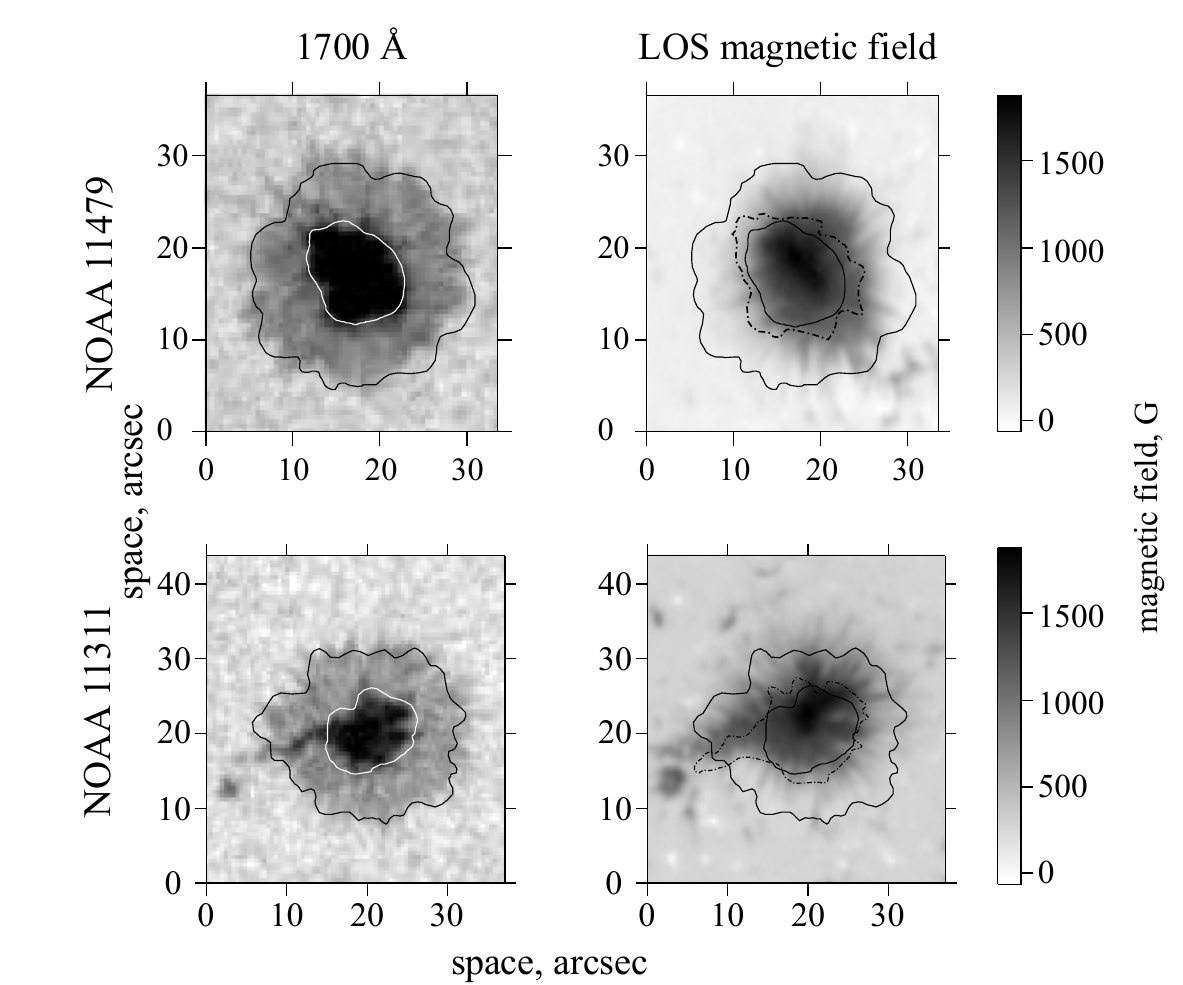}
}
\caption{Sunspots under investigation. Sunspot images
in the 1700\,\AA\ band (left), magnetograms (right). The inner and outer penumbra
boundaries (black and white solid lines) are outlined by the 0.1 and 0.5 quiet Sun 1700\,\AA\ band intensity levels. 
Zero level corresponded to the lowest umbral intensity.
The dashed isolines mark the regions where the magnetic field inclines at 45$\degree$
to the surface normal. }
\label{fig:ao-boundaries}
\end{figure}

\section{Results} 
      \label{S-general}
\subsection{Oscillations and Fan Structures Above Sunspots}
\par Active regions NOAA 11311 and 11479 are single, round, medium-sized sunspots 
(see Figure\,\ref{fig:ao-boundaries}).
We observed them near the central meridian, which
allowed us to minimize the projection effect on the analysis of oscillations at
different heights. The domains studied included sunspots and the neighboring
regions up to 5\,--\,10\arcsec\ from the outer penumbra borders, since
sunspots affect the oscillation characteristics of these regions \citep{kob2000sp}.


\begin{figure}
\centerline{
\includegraphics[width=11cm]{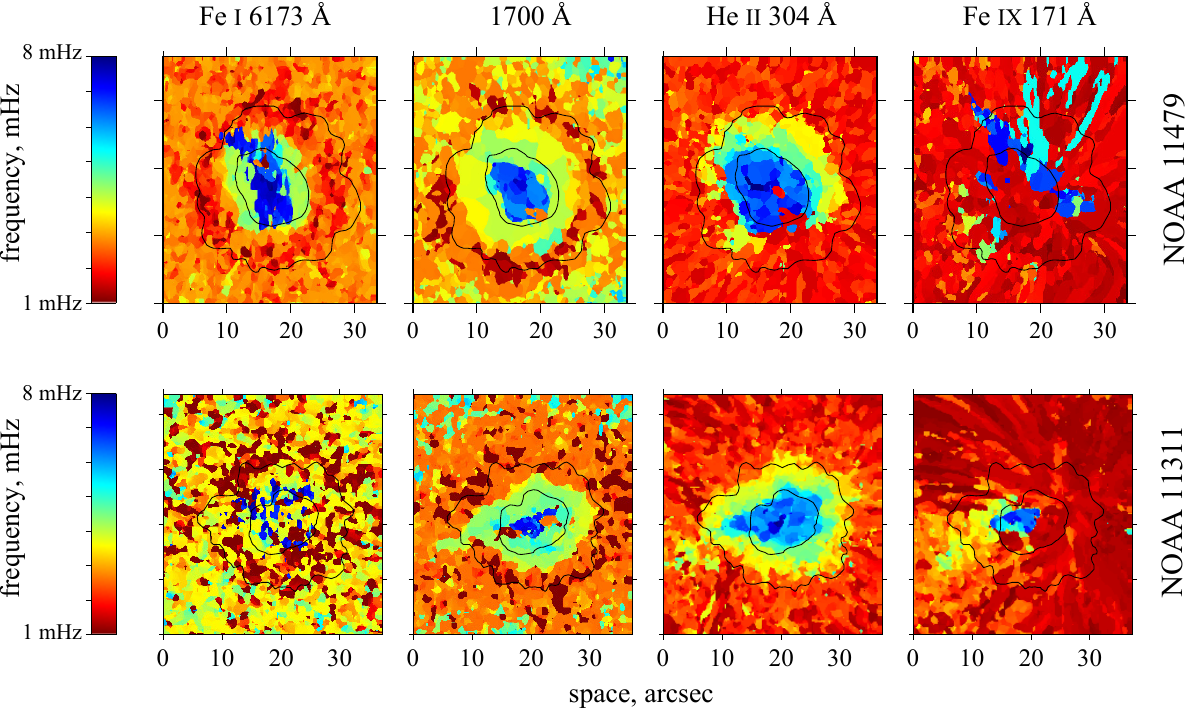}
}
\caption{Spatial distributions of dominant frequencies from the photosphere to
the corona. The black closed lines mark the inner and outer penumbral
boundaries.}
\label{fig:osc-1-8-mhz}
\end{figure}

\par
\begin{figure}
\centerline{
\includegraphics[width=11cm]{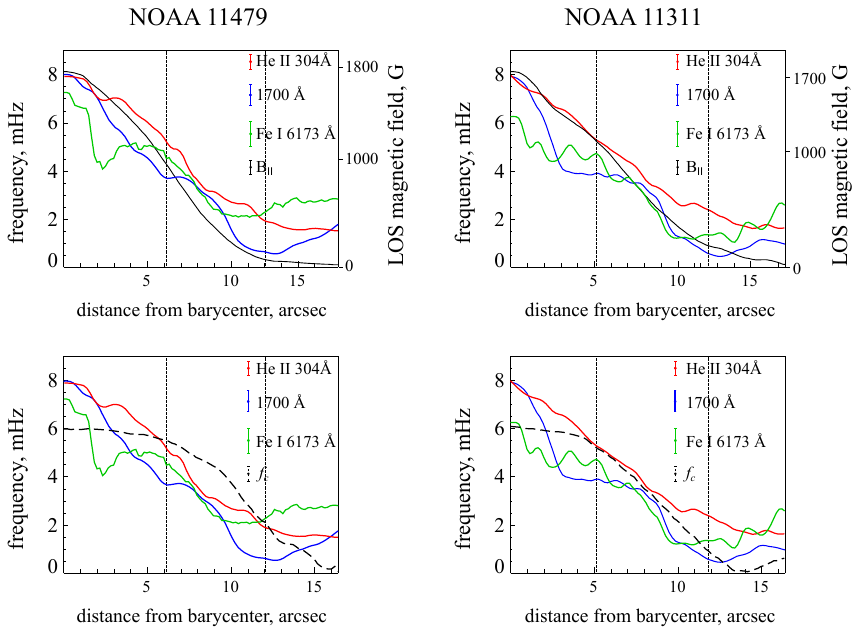}
}
\caption{Dominant frequency as a function of distance from sunspot barycenters [$F(r)$].
Green line is Fe\,\textsc{i} 6173\,\AA; blue, 1700\,\AA; red, 304\,\AA. Black solid lines in
the top panels represent LOS magnetic field as a function of distance
from sunspot barycenters. The dashed lines in the bottom panels represent
cut-off frequency as a function of distance from sunspot barycenters deduced
using Equation\,(\ref{eq:fc}) for acoustic speed of 6.2\,km\,s${^{-1}}$ at 
the 6173\,\AA\ line formation level. The vertical lines mark the inner and outer penumbral boundaries.}
\label{fig:freq-distance}
\end{figure}

\begin{figure}
\centerline{
\includegraphics[width=11cm]{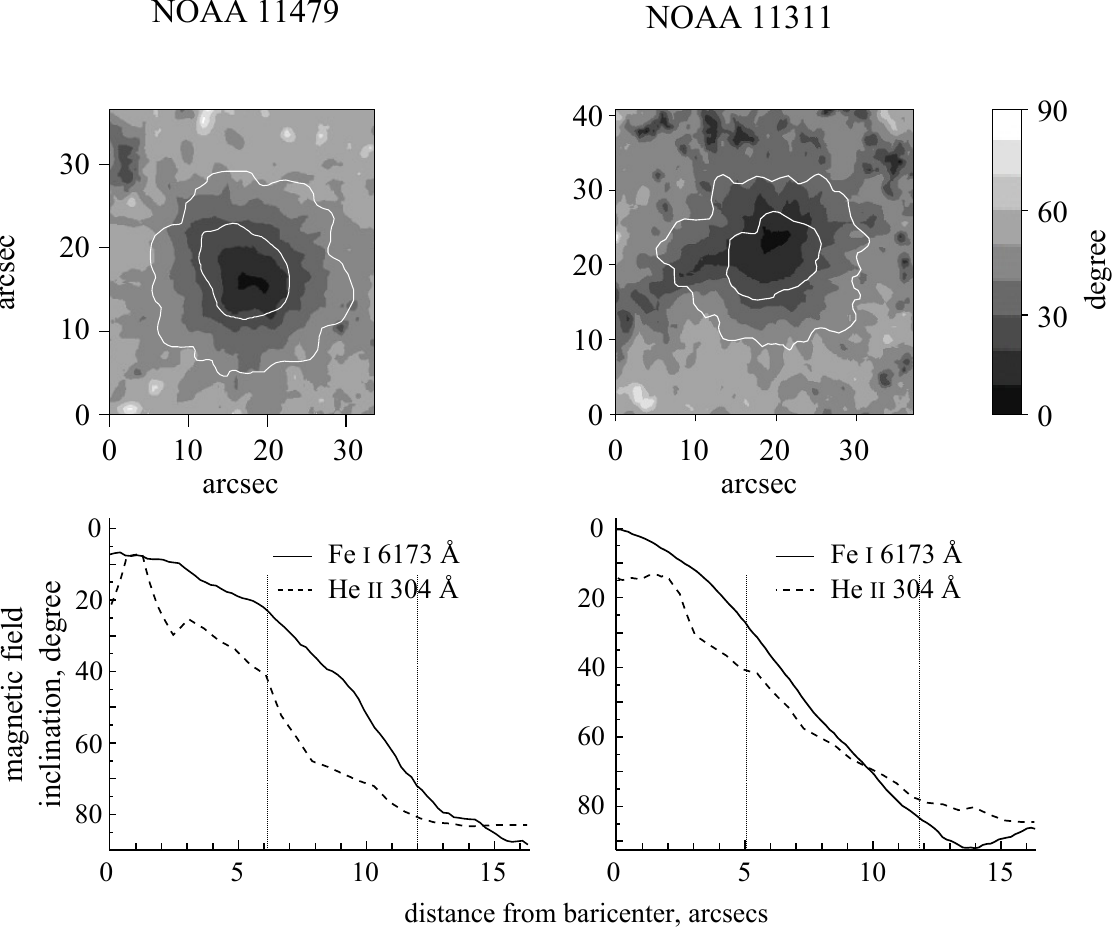}
}
\caption{Magnetic-field inclination angle to solar normal gray-scale maps (upper row). 
The white closed lines mark the inner and outer boundaries of the penumbrae. 
Magnetic-field inclination angle to solar normal as a function of distance from sunspot
barycenters (bottom row). The solid line was derived from vector magnetic-field measurements
for the Fe\,\textsc{i} 6173\,\AA\ level (SDO/HMI). The dashed line was obtained using
Equation\,(\ref{eq:fi}) for the He\,\textsc{ii} 304\,\AA\ level. The thin vertical lines mark the
inner and outer penumbral boundaries.}
\label{fig:mf-incl-distance}
\end{figure}

 Figure\,\ref{fig:osc-1-8-mhz} shows the dominant frequency for every
spatial element of the sunspots at several heights: Fe\,\textsc{i} 6173\,\AA,
1700\,\AA, He\,\textsc{ii} 304\,\AA, and Fe\,\textsc{ix} 171\,\AA\ bands. The frequency
determination was performed as follows.
Firstly, the FFT intensity spectrum for each 0.6\arcsec$\times$0.6\arcsec domain
was obtained. Preparing the signals for the FFT analysis, we subtracted an average intensity 
and applied a $\mbox{cos}^{2}$ bell filter to minimize effects of abrupt endings. Then, the integral power in a 1\,mHz rectangular window
was calculated. The window was moved throughout the spectrum with a 0.02\,mHz step
giving a set of values. The frequency corresponding to the maximum value was
taken as the dominant one for the particular spatial domain. This representation
is a convenient way to obtain a picture of the spatial distribution of oscillation frequencies 
above the sunspots under study. Five-minute oscillations dominate
in the lower photosphere (the Fe\,\textsc{i} 6173\,\AA\ line), forming typical
fragmented structure in the umbrae and the neighboring outer penumbra regions.
Photospheric three-minute oscillations in sunspot NOAA 11311 are located mainly
along the umbra boundary instead of the umbral central part,
where these oscillations are located in the
1700\,\AA, He\,\textsc{ii} 304\,\AA, and Fe\,\textsc{ix} 171\,\AA\
bands. This corresponds to our earlier results \citep{kobanov2011,kob2013aa}.
Five-minute oscillations are located in an annular zone around the sunspot center, which expands beyond the umbra boundaries with increasing
height in both sunspots.
Low-frequency oscillations start in the lower photosphere of the penumbra
(left panels in Figure\,\ref{fig:osc-1-8-mhz}). Such a topology of the
frequency distribution height cuts supports the concept explaining
RPW as a ``visual pattern''
\citep{rouppe2003,bloomfield2007a,kobkol2009}. In the upper layers of
the solar atmosphere, the field of view is mostly occupied by the low-frequency
oscillations marked in red (Figure\,\ref{fig:osc-1-8-mhz}),
whereas the high-frequency oscillations marked with blue color reside within
the inner-penumbra boundaries. The presence of the low-frequency inclusions
inside the umbrae can probably be explained by fine-structure
irregularities of the magnetic field. One should note that Figure\,\ref{fig:osc-1-8-mhz} does not
give the real picture of the oscillation power; it merely shows the
spatial distribution of the frequencies instead. Circular-shaped frequency spatial
localization allowed us to plot the frequency \textit{vs.} distance from barycenter
for three height levels: Fe\,\textsc{i} 6173\,\AA, 1700\,\AA, He\,\textsc{ii} 304\,\AA\
(see Figure\,\ref{fig:freq-distance}).
White-light images were used to find barycenters of the sunspots.
Then, using the data in Figure\,\ref{fig:osc-1-8-mhz}, we calculated a mean frequency value [$F(r)$]
averaged over the points located at distance
$r$ from the barycenter (Figure\,\ref{fig:freq-distance}). Each point of the curves
was determined as described. The radial spatial step was chosen to be 0.6\arcsec.
Similarly, the longitudinal magnetic field [$B_{\shortparallel}$] in Figure\,\ref{fig:freq-distance} and the absolute
value of the magnetic-field inclination in Figure\,\ref{fig:mf-incl-distance} (solid lines) were plotted against the distance from the
barycenter.
Magnetic-field inclination angle [$\varphi$] to the solar-surface normal was
calculated using SDO/HMI data as described by \cite{borrero2011sp} and converted following the formalism of \cite{GaryHag1990}.
It is of interest to compare the derived dependencies with the
cut-off frequency for the waves, using the slow magnetoacoustic wave
approximation. For this purpose, we used the following equation \citep{McinJef2006,Botha2011}:
\begin{equation}
f_c=\frac{g_0\gamma \mbox{cos}\varphi(r)}{4\pi v_s}
\label{eq:fc}
\end{equation}
where $f_c$ is the cut-off frequency; $g_0$=274 $\mbox{m\,s}^{-2}$ is the gravitational constant;
$v_s$ is the speed of sound; $\gamma=\frac{5}{3}$ is the
adiabatic index; $\varphi(r)$ is the dependence of inclination angle on
the barycenter distance. Finally, we obtain the dashed curves in
Figure\,\ref{fig:freq-distance} expressing the dependence of $f_c$ on
the barycenter distance for both sunspots.

\par Analyzing the plots in Figure~\ref{fig:freq-distance},
one can see several features: curves of the lowest atmospheric layers
(Fe\,\textsc{i} 6173\,\AA\ and 1700\,\AA) show a jump. This is typical
for both sunspots and is thus unlikely to be an artifact. The curves
expressing frequency \textit{vs.} distance from the barycenter
do not coincide with those for $f_c$ calculated using Equation~(\ref{eq:fc}).
This is especially evident in the umbra, where the $f_c$ curve is almost horizontal.
Note that more correspondence is found for the $B_{\shortparallel}$
curve in Figure\,\ref{fig:freq-distance} (upper panels).
All of the curves approach each other in the middle parts of the
penumbrae. This implies that the
3\,--\,3.5\,mHz oscillations observed in the corresponding penumbra regions prevail
at these heights. Probably, this circular penumbral zone is the very place
to search for the upwardly propagating five-minute waves.
\par As \cite{jess2013} showed, the dominant frequency distribution can be
used to study physical conditions at corresponding heights (\textit{e.g.} the
magnetic-field inclination). Substituting $f_c$ with
$F(r)$ measured at the He\,\textsc{ii} 304\,\AA\ line formation level in Equation~(\ref{eq:fc}), one can
roughly estimate the magnetic-field inclination at this height:
\begin{equation}
\varphi = \mbox{arccos}\left(\frac{4\pi v_s F(r)}{g_0\gamma} \right)
\label{eq:fi}
\end{equation}
Direct substitution of the dominant frequency [$f_c$] for cut-off frequency [$F(r)$] results in 
the ratio in the parentheses being greater than 1 for the high frequencies, which 
is unacceptable. Following \cite{yuan2014aa}, we used an empirical ratio $f_c=$0.81F(r)\,mHz$-1$\,mHz 
for the 304\,\AA\ line. The result is suitable for qualitative assessments. To acquire more 
precise quantitative assessments the empirical ratio between $f_c$ and $F(r)$ should be 
determined based on greater statistics.
The results of such an estimation are presented in Figure~\ref{fig:mf-incl-distance},
where the solid lines denote the magnetic field inclination angles derived with
SDO/HMI in the photospheric Fe\,\textsc{i} 6173\,\AA\ line, and the dashed lines
denote those derived using Equation~(\ref{eq:fi}) for the
He\,\textsc{ii} 304\,\AA\ line. The inclination reaches \textbf{40\textdegree\ }
at the inner penumbra boundary. Close values
were obtained by other authors \citep{jess2012apj,rez2012ApJ,kob2013aa},
who used different methods. In accordance with the derived
plots, the magnetic-field inclination angle at the
He\,\textsc{ii} 304\,\AA\ line formation level rapidly increases in the umbrae and the
adjacent penumbrae, and this rate gradually decreases in the outer penumbrae.
This dependence corresponds to the plot presented by \cite{yuan2014aa}
for the He\,\textsc{ii} 304\,\AA\ line.

\begin{figure}
\centerline{
\includegraphics[width=11cm]{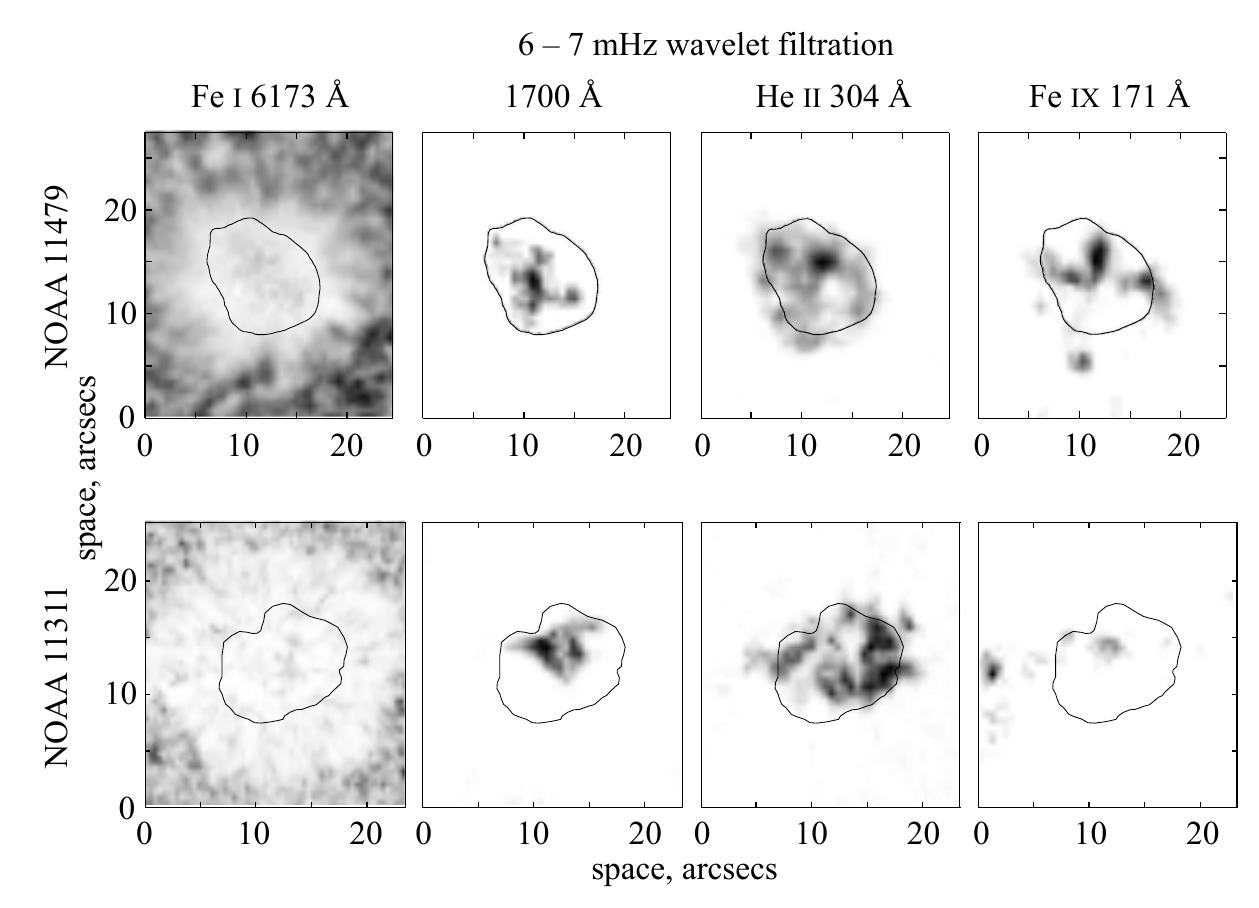}
}
\caption{Spatial localization of high-frequency oscillations (6\,--\,7\,mHz) at
the different atmosphere levels. Darker regions correspond to more powerful oscillations.
Solid lines mark the umbral boundaries.}
\label{fig:6-7-mhz-wavelet-2d}
\end{figure}

\begin{figure}
\centerline{
\includegraphics[width=12.5cm]{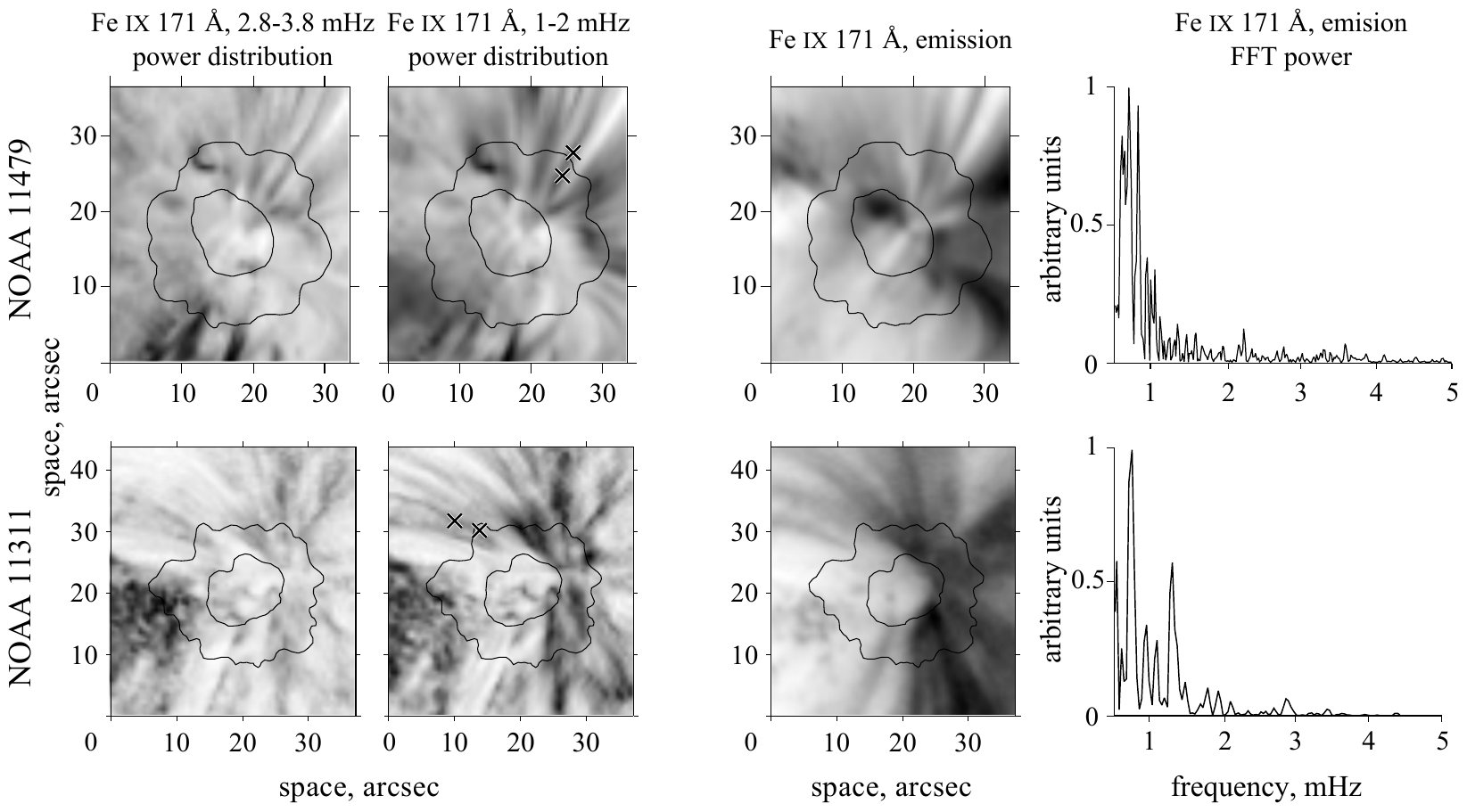}
}
\caption{Fan structures in spatial localizations of low-frequency coronal
oscillations (columns\,1 and 2). Inverse sunspot images in the 171\,\AA\ line intensity (column 3). Intensity oscillation power spectra of the 171\,\AA\ line averaged
over the fields of view (column 4).}
\label{fig:fan-wavelet-and-171-int}
\end{figure}

\par As was noted above, the low-frequency oscillations
occupy a greater area in the FOV with increasing height (Figure\,\ref{fig:osc-1-8-mhz}).
At the same time, image fragments in the power maps become elongated
in the radial direction. Frequency filtering provides clear
visualization of spatial localization of different modes.
Figure\,\ref{fig:6-7-mhz-wavelet-2d} presents 6\,--\,7\,mHz filtered intensity signals corresponding to
several heights. High-frequency oscillation locations at all levels
are confined to the domains of the photospheric umbral boundaries. No fan
structures are seen in this image. The situation is different for
low-frequency oscillations of the 3--3.8 and 1--2 mHz bands.
We refer to Figure\,\ref{fig:fan-wavelet-and-171-int}, as we are most interested
in finding the wave frequencies that penetrate into the corona and dominate
in the fan structures at the Fe\,\textsc{ix} 171\AA\ line level.
Left and middle panels in this figure show spatial distribution of
the oscillation power in the 3\,--\,4 and 1\,--\,2\,mHz bands respectively; right panels
show the Fe\,\textsc{ix} 171\AA\ intensity images, averaged over the
entire time series. The intensity images were inverted for the
convenience of visual comparison. One can see that coronal-fan
structures are best reproduced in the middle panels showing the
1--2 mHz band oscillation power. Joint analysis of Figures\,\ref{fig:fan-wavelet-and-171-int}
and \ref{fig:osc-1-8-mhz} leads to the following conclusions:
i) the 10\,--\,15 minute oscillations dominate in the corona above
sunspots; ii) these oscillations are observed along the
loops;
thus, one can trace changes in their properties
along the horizontal part of the loops and try to measure
the propagation speed.

\begin{figure}
\centerline{
\includegraphics[width=11cm]{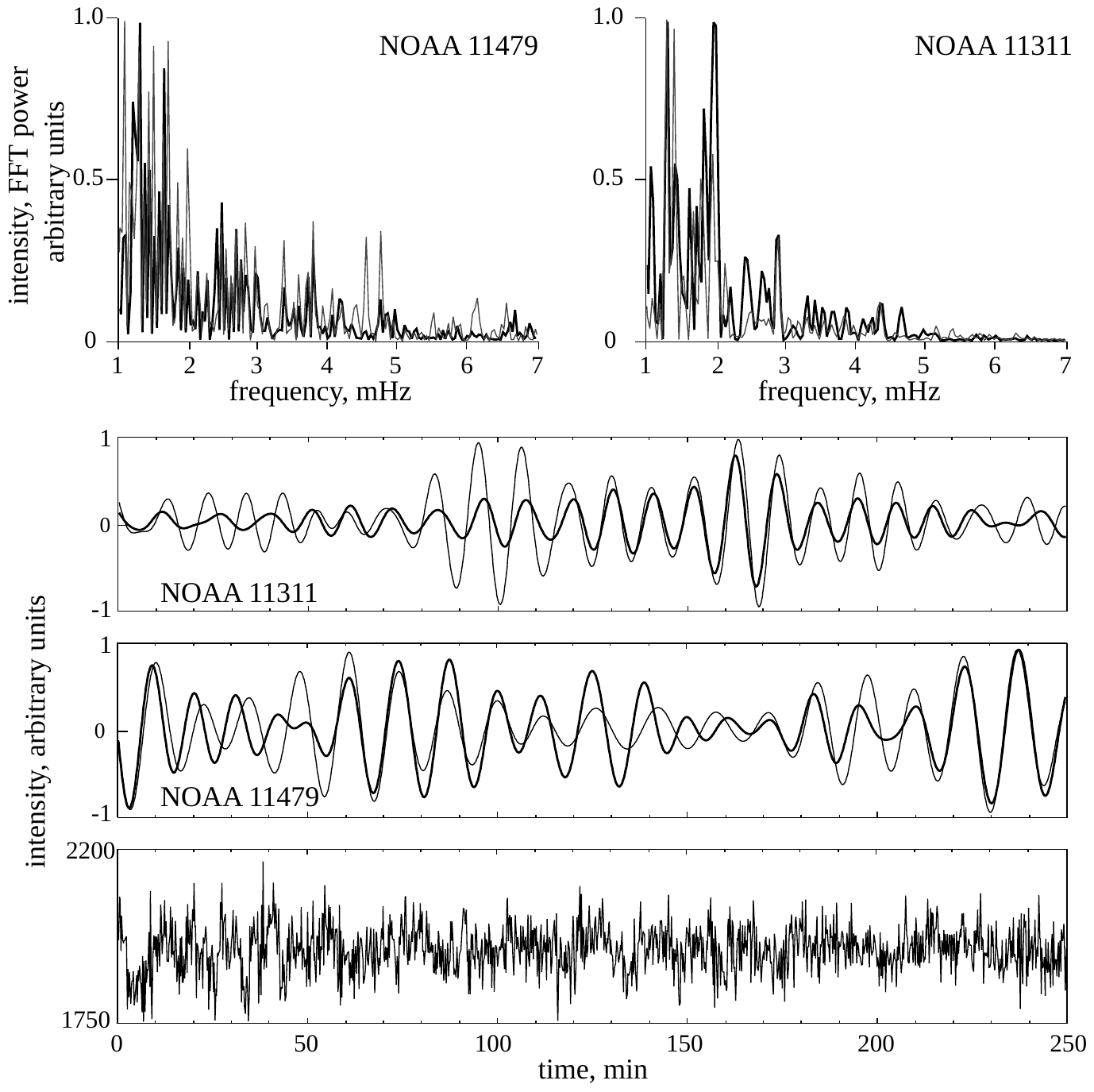}
}
\caption{Oscillation behavior in coronal loops above sunspots.
The top panels show intensity oscillation power spectra in the loop points
marked with crosses in Figure\,\ref{fig:fan-wavelet-and-171-int}. The results
of frequency filtration are presented in the middle panels. The thick
lines show spectra and signals at the points located further from the sunspot centers. The bottom panel shows an example of an original intensity signal (NOAA\,11479).}
\label{fig:spectral-phase}
\end{figure}

\par The crosses in the middle panel of Figure\,\ref{fig:fan-wavelet-and-171-int} show the points
that were analyzed in detail.
The power spectra for these points are shown in Figure\,\ref{fig:spectral-phase}.
A correspondence between the spectra
indicates that both of them belong to the same loop.
Contrary to our expectations, the time delay between the
signals from two points of the same loop was ambiguous, with variations
in amplitude and sign over the whole time series.
One of the possible explanations implies the existence of several thin
and highly transparent coronal loops
contributing to the signal.
In spite of such a phase difference, the oscillations with close frequencies
have similar power spectra. The averaged estimations give the following
phase velocities: 120\,$\pm$\,25\,km\,s$^{-1}$ for NOAA\,11479 and
130\,$\pm$\,30\,km\,s$^{-1}$ for NOAA\,11311. Similar values were obtained
by \cite{Nightingale1999,Robbrecht2001,MarshWalsh2003}.

\subsection{Characteristics of Facular Oscillations}
\par The analysis was performed using two faculae. Facula No.\,1
was observed on 6 October 2011 from 00:47\,UT untill 02:42\,UT, with
center coordinates S12\,E05. This facula is of special interest due to
its proximity to sunspot NOAA\,11311, which implies its direct connection
through magnetic field arches. We observed facula region No.\,2
with coordinates N13\,W08 on 1 October 2011, from
03:41\,UT untill 05:06\,UT. We identified this facula as not being
connected with sunspots. The ground-based observations are spectrogram
time series in the Si\,\textsc{i}\,10827\AA\ and He\,\textsc{i}\,10830\AA\
lines. The temporal resolution was 3.3\,seconds, and the spatial resolution along the slit
was 1\,--\,1.5\arcsec\ on average. We chose the same SDO lines that we used for the sunspot analysis.
Facular regions were confined by the 0.7 brightness isoline in the 1700\,\AA\ band,
where the 100\,\% brightness was considered to be the maximum facula-core brightness. The faculae have arbitrary shapes and lack radial symmetry.
 We intended to determine
the frequency bands which prevail in the fan structures observed at
the 171\,\AA\ line formation level, and to reveal their source as far as
possible. To this end, we needed to study facular
oscillation properties from the deep photosphere to the corona.

\begin{figure}
\centerline{
\includegraphics[width=10.5cm]{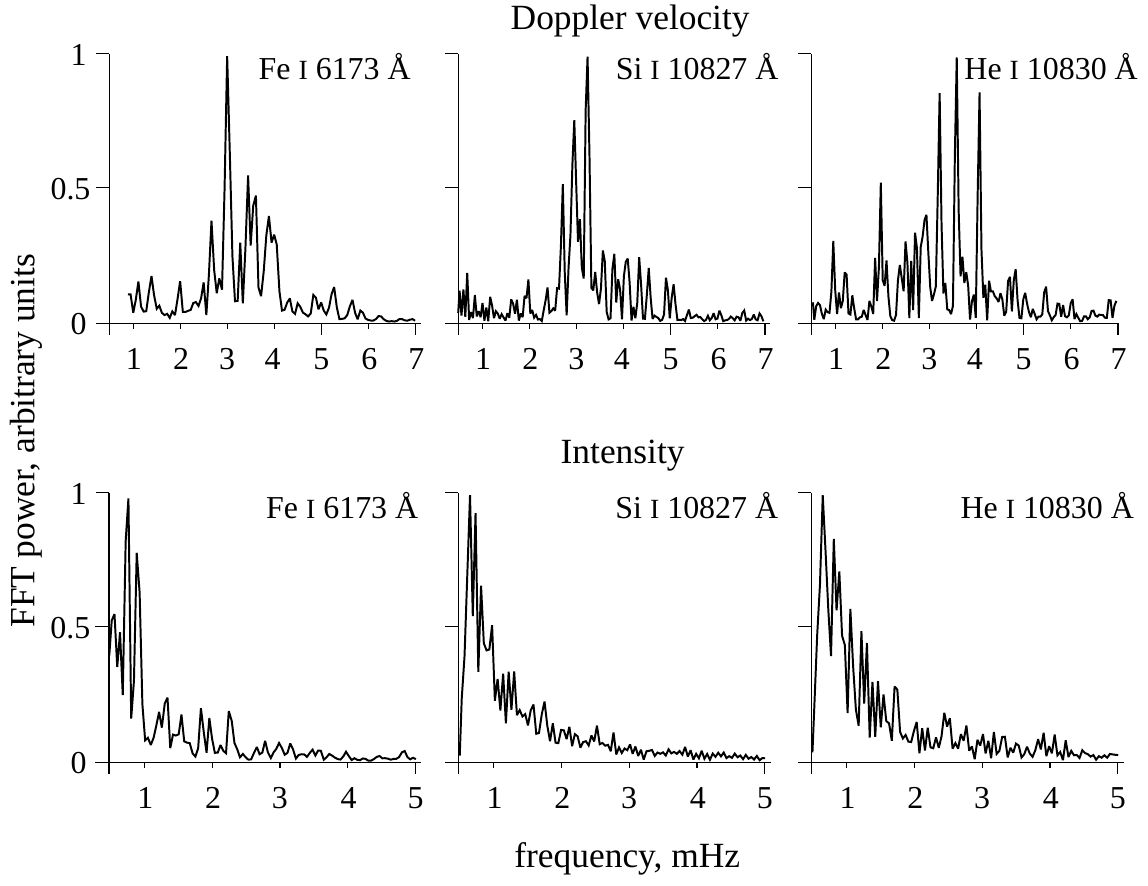}
}
\caption{Power spectra of the LOS velocity (top panels) and intensity (bottom
panels) oscillations. The signals were averaged over the area marked with the
square box in Figure\,\ref{fig:fac2-wavelet-2d} (first column).}
\label{fig:fft-velocity-intensity}
\end{figure}

\begin{figure}
\centerline{
\includegraphics[width=11.5cm]{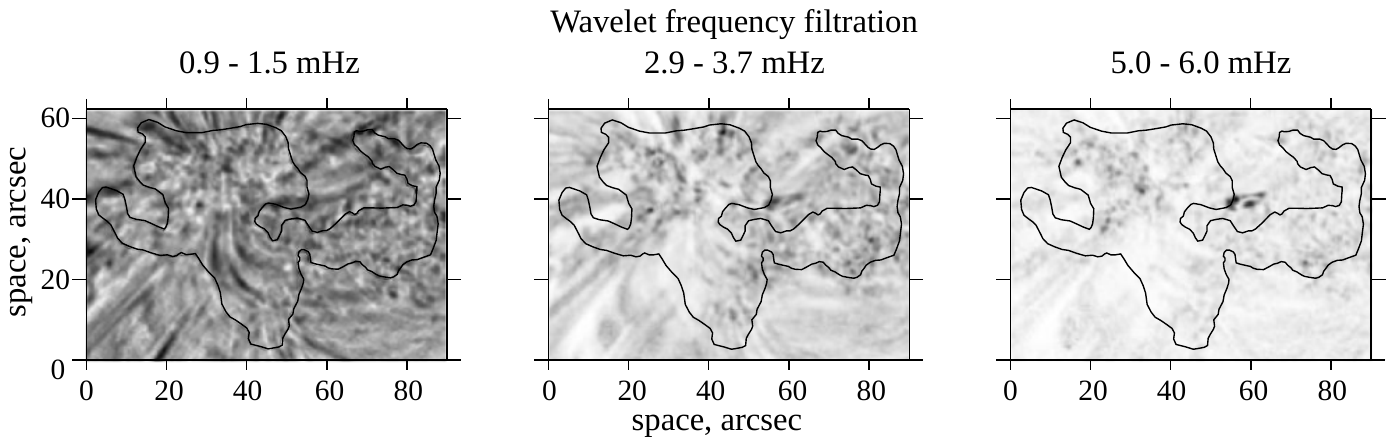}
}
\caption{Spatial localization of the Fe\,\textsc{ix}\,171\AA\ line intensity oscillations
for three frequency ranges.
Darker regions correspond to more powerful oscillations. Solid lines mark the facular boundaries (0.7 brightness
isoline in the 1700\AA\ band).
}
\label{fig:171-wavelet}
\end{figure}

Information about oscillation parameters in the upper atmosphere of faculae is controversial.
\cite{1983A&A...120..185K} observed periodic displacements of the coronal 5303\,\AA\
line at a height of 25\,000\,--\,30\,000\,km above a facula. They revealed 300\,seconds, 80\,seconds, and
40\,seconds periods. Using high spatial resolution
observations, \cite{dewijn2009} found that three-minute periods prevail in the chromosphere
above the facula core, while five-minute periods dominate on its periphery. This configuration is similar to that observed
in sunspots. When studying five-minute oscillations in the
X-rays above faculae, \cite{2011ApJ...738L...7D} concluded that they were related to global solar surface
oscillations ($p$-modes) observed in the photosphere. \cite{ofman1999ApJ} and
\cite{deforest1998ApJ} recorded 15-minute oscillations in the corona above polar
faculae, which they explained as a manifestation of slow magnetoacoustic waves.

\par First, we analyzed characteristics of LOS-velocity oscillations
observed at three heights: Fe\,\textsc{i}\,6173\AA\ -- 200\,km
\citep{parnell1969sp}, Si\,\textsc{i}\,10827\AA\ -- 540\,km;
He\,\textsc{i}\,10830\AA\ -- 2000\,km \citep{centeno09}.

\begin{figure}
\centerline{
\includegraphics[width=11cm]{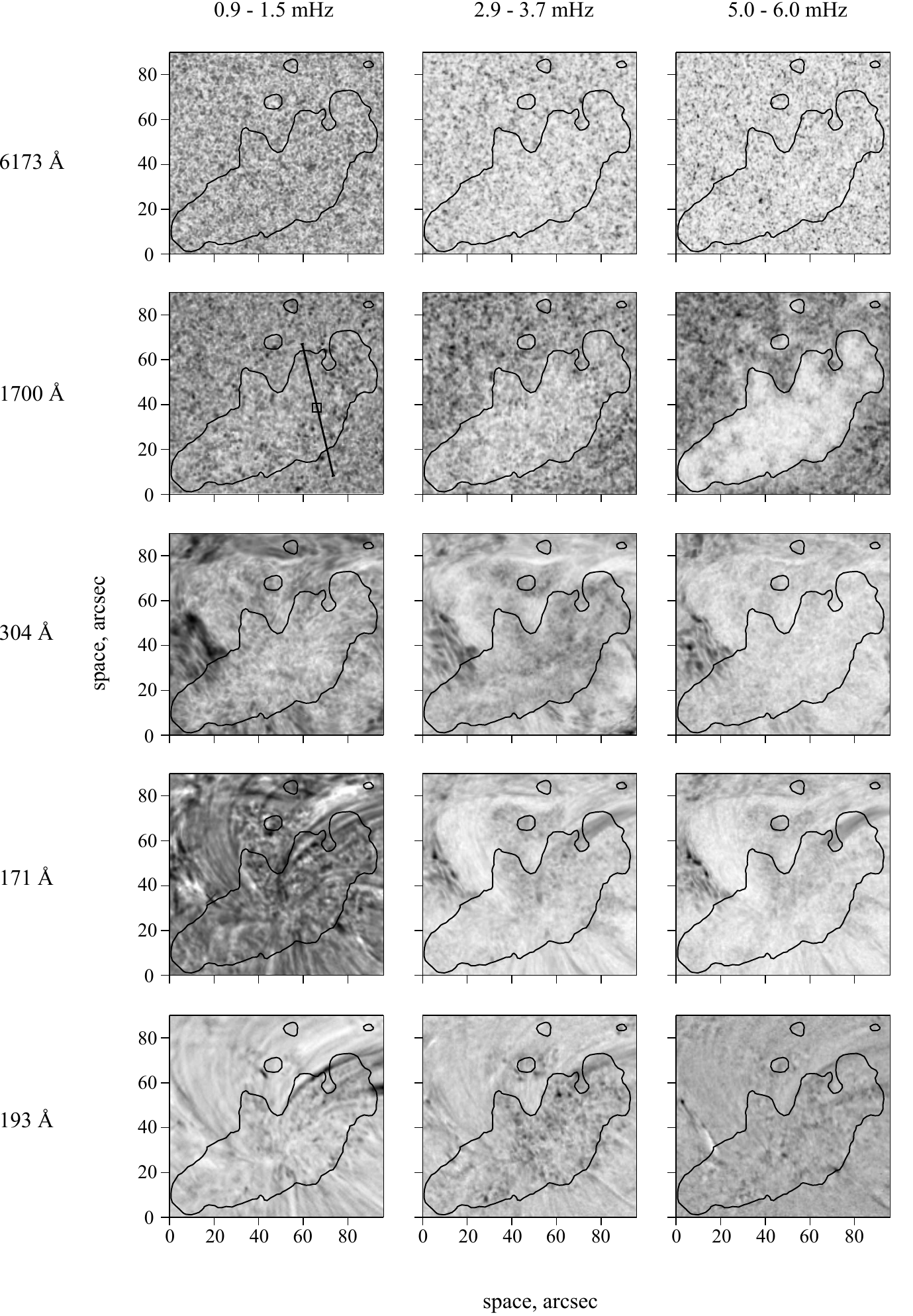}
}
\caption{Spatial distribution of oscillation power above Faculae\,2 in the three
frequency ranges for five atmosphere levels. Darker regions correspond to more powerful oscillations.
Solid lines mark the facular boundaries.}
\label{fig:fac2-wavelet-2d}
\end{figure}

\par The observed LOS-velocity signals contain information about acoustic oscillations
at these heights. Figure\,\ref{fig:fft-velocity-intensity} shows the power spectra for the signals averaged over the
area marked with a square in the middle part of Facula 2 in Figure\,\ref{fig:fac2-wavelet-2d} (first
column, second panel). Both space (Fe\,\textsc{i}\,6173\AA) and ground-based
(Si\,\textsc{i}\,10827\AA, He\,\textsc{i}\,10830\,\AA) observations of LOS velocity show that the
five-minute period is dominant in the photosphere and upper chromosphere
(Figure\,\ref{fig:fft-velocity-intensity}, top row). The lower row in Figure\,\ref{fig:fft-velocity-intensity} presents the intensity
variation spectra in the same locations. All spectra are normalized to their
maximum values. The five-minute oscillations are almost absent in the intensity
spectra, while they dominate in the LOS-velocity spectra. The 0.7\,--\,1.2\,mHz oscillations
dominate in the intensity spectra of the lower photosphere. There have always
been doubts about the solar origin of intensity oscillations in ground-based observations: are they a result of the
Earth's atmospheric turbulence? The photospheric spectra of space
observations (Fe\,\textsc{i}\,6173\,\AA) agree well with those of
ground-based observations (Si\,\textsc{i}\,10827\,\AA). The height difference
between the formation levels (200\,km and 540\,km for the Fe\,\textsc{i} and
Si\,\textsc{i} lines respectively) may explain the minor difference in
details. At the same time, the photospheric and chromospheric ground-based
spectra also differ, which should not be the case if they are affected by
the same artifact. These arguments confirm the solar
origin of the low-frequency peak shown in the lower panels in Figure\,\ref{fig:fft-velocity-intensity}.
The absence of such peaks in the LOS-velocity variation spectra may signify
that two different types of oscillations coexist in the observed volume.
The five-minute variations in the LOS-velocity signals are undoubtedly
acoustic oscillations, and the 0.7\,--\,1.2\,mHz peaks are probably a
sausage-mode manifestation.

\par
\begin{figure}
\centerline{
\includegraphics[width=8.1cm]{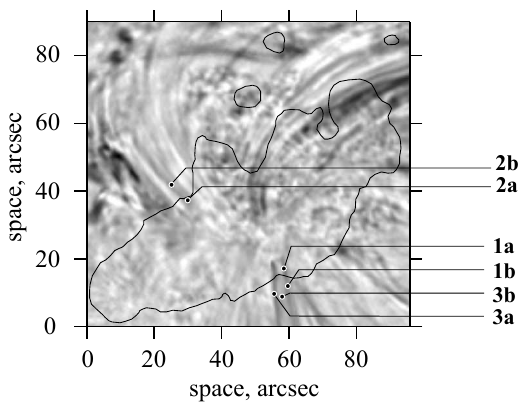}
}
\caption{Manifestation of fan structures in spatial distribution of low-frequency coronal oscillations
at the Fe\,\textsc{ix}\,171\AA\ line-formation height. The marks show the
loop element pairs selected for the analysis.}
\label{fig:fan-low-freq}
\end{figure}

\begin{figure}
\centerline{
\includegraphics[width=10cm]{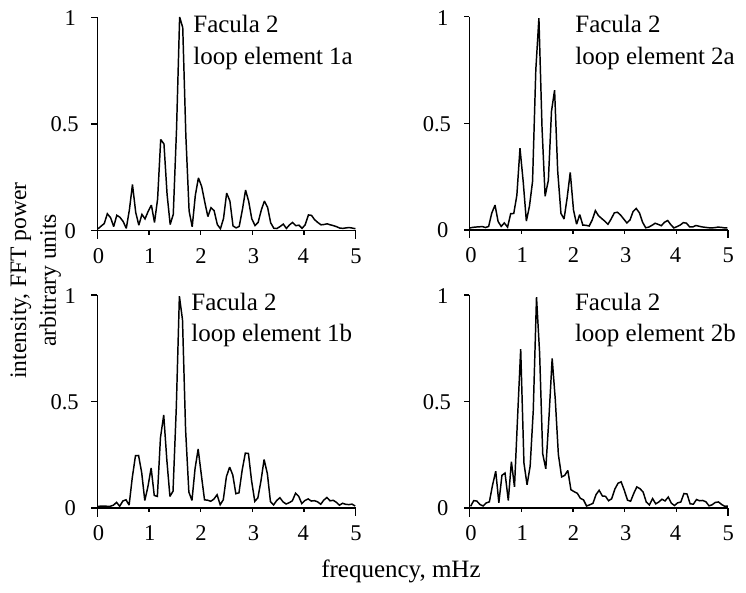}
}
\caption{Oscillation power spectra for the loop elements marked in Figure\,\ref{fig:fan-low-freq}.}
\label{fig:fft-loop-low-freq1}
\end{figure}

\begin{figure}
\centerline{
\includegraphics[width=5cm]{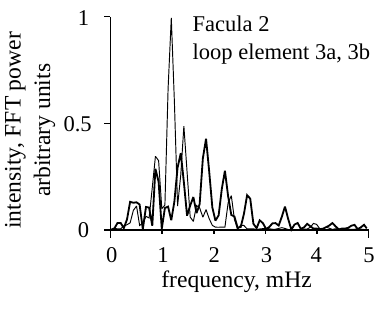}
}
\caption{Oscillation power spectra for the points located inside (thin line) and
outside (thick line) the loop.}
\label{fig:fft-loop-low-freq-2-in-out}
\end{figure}

 Figure\,\ref{fig:171-wavelet} shows how modes with different frequencies are distributed in the lower corona
(the Fe\,\textsc{ix}\,171\,\AA\ line). Spatial distributions of low-frequency oscillation power
outlines fan-loop structures best. This similarity is less
pronounced in the five-minute range and almost disappears in the three-minute range.
The change in power spatial localization of these frequencies with height
can be illustrated by the example of Facula~2 (Figure\,\ref{fig:fac2-wavelet-2d}). At the lowest
level (Fe\,\textsc{i} 6173\,\AA), the facula region is similar to the surrounding
background at all frequencies (see Figure\,\ref{fig:fac2-wavelet-2d}, upper row). At the
temperature minimum level (the 1700\,\AA\ line) the decrease in oscillation
power becomes evident, relative to the surrounding background. This is most
apparent at high frequencies (Figure\,\ref{fig:fac2-wavelet-2d}, the third panel in the second row). The oscillation-power distribution in the
transition region (the He\,\textsc{ii}\,304\,\AA\ line) shows the first signs of
elongated elements, which become a clearly expressed fan structure in the lower
corona (Fe\,\textsc{ix}\,171\,\AA) in the frequency range of 0.9\,--\,1.5\,mHz. This structure 
has reduced contrast at the Fe\,\textsc{xii, xxiv}\,193\,\AA\ line formation height. At the level of the Fe\,\textsc{xii, xxiv}\,193\,\AA\ line,
the facula shows higher oscillation power relative to the adjacent
regions at the highest level in the 2.8\,--\,3.8\,mHz and 5.0\,--\,6.0\,mHz ranges.
Probably, this tendency is better developed higher above the Fe\,\textsc{xii, xxiv}\,193\,\AA\ line
formation height. This will be a subject for further research. The fact that
spatial localization of low-frequency oscillations closely reproduces coronal-loop structures
means that facular low-frequency oscillations penetrate to the corona through
this very loop system. We observe the upper and, thus, the most horizontal parts
of the loops at the Fe\,\textsc{ix}\,171\,\AA\ line-formation level. Such a feature of the
loop geometry gives an opportunity to analyze the oscillation spectral-phase
characteristics in more detail. The marks in Figure\,\ref{fig:fan-low-freq} show points that we used
to calculate the FFT power spectra. Figure\,\ref{fig:fft-loop-low-freq1} represents the spectra for each element pair
(1.2${\times}$1.2\arcsec\ in size) located at individual loop. These spectra
are convincing evidence of the fact that low frequencies dominate in loop structures at the 171\,\AA\ line
formation height. May these oscillations exist outside loops as well?
Figure\,\ref{fig:fft-loop-low-freq-2-in-out} gives answers to this question, showing the spectra for two
spatial elements, one inside the loop and the other one ${\approx}$2\arcsec away
from the loop. The oscillation inside the loop element is approximately five times
higher than the background level. Spectral similarity for the loop elements gives
hope that we are able to unambiguously measure the phase speed at the dominant
frequency. However, the results of a similar analysis were ambiguous
\citep{kob2014ARep}. The phase speed based on the time-lag measurement varies
from 50\,km\,s$^{-1}$ to 100\,km\,s$^{-1}$ with the average value of 60\,km\,s$^{-1}$.


\begin{figure}
\centerline{
\includegraphics[width=9.5cm]{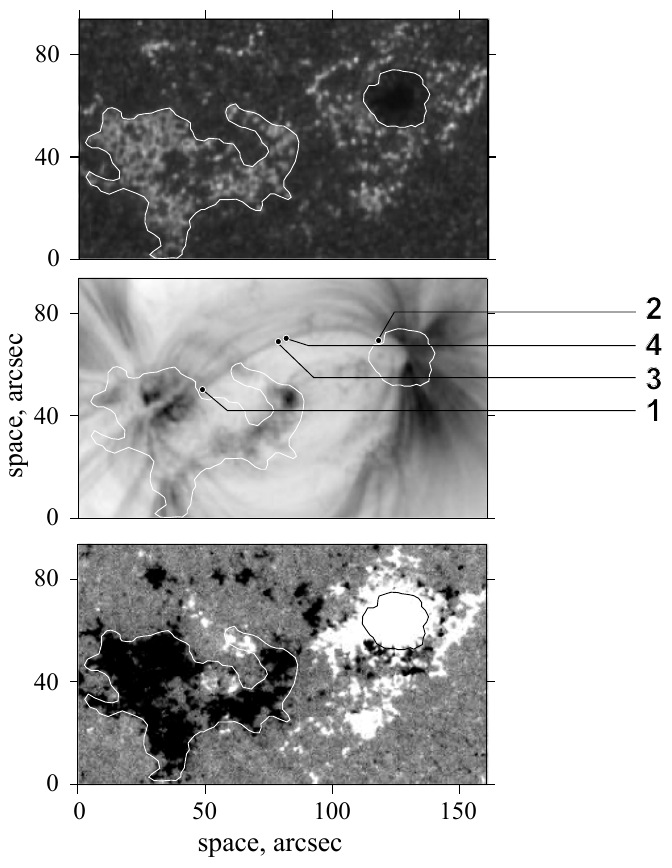}
}
\caption{Sunspot NOAA\,11311 and the adjacent facula. Top panel: the
1700\AA\ band image; middle panel: 1--2 mHz oscillation wavelet power
in the Fe\,\textsc{ix}\,171\AA\ line; bottom panel: LOS magnetogram.
Darker regions correspond to more powerful oscillations.
Solid lines mark the facular and sunspot's boundaries.
}
\label{fig:sunspot-fac-171-2d}
\end{figure}

\par
\begin{figure}
\centerline{
\includegraphics[width=10cm]{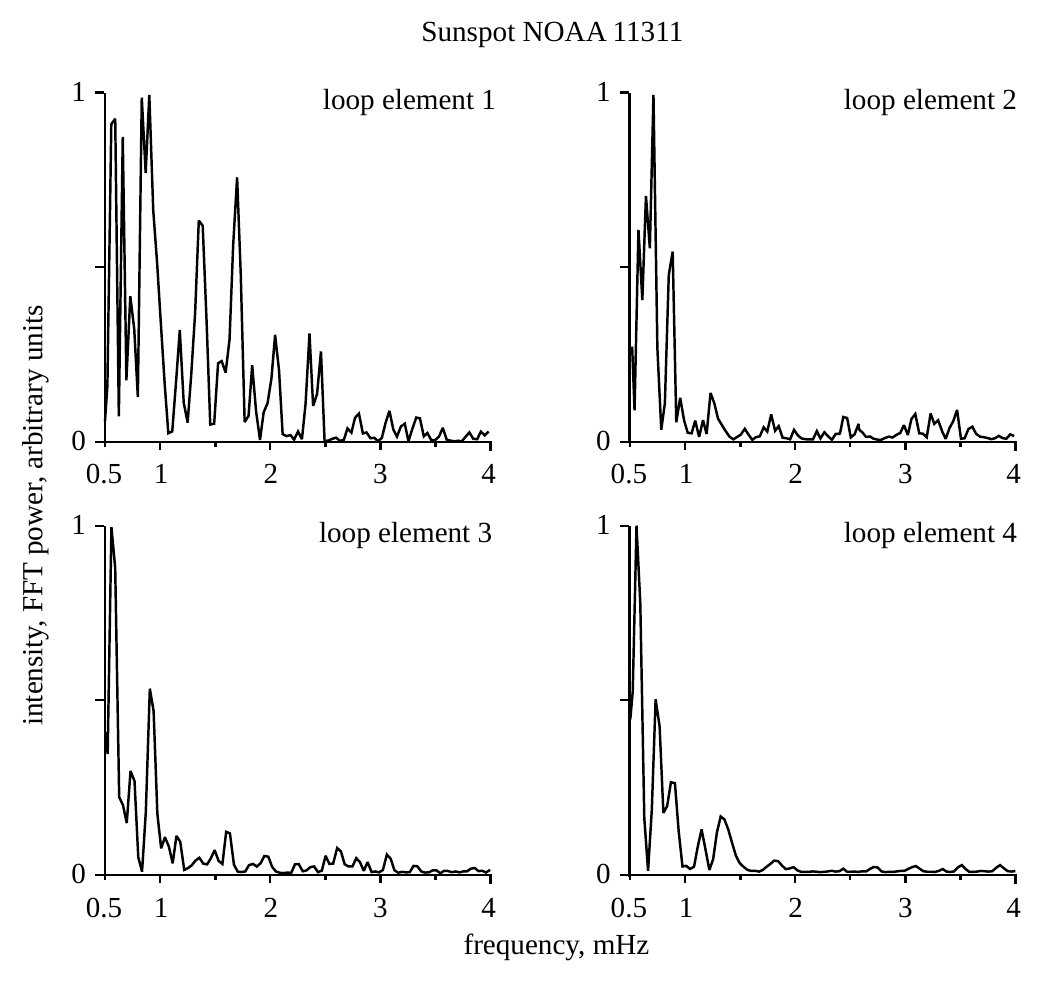}
}
\caption{FFT power spectra. Oscillations detected at the points marked in
Figure\,\ref{fig:sunspot-fac-171-2d} (middle panel).}
\label{fig:fft-11311-4-points}
\end{figure}

\begin{figure}
\centerline{
\includegraphics[width=11cm]{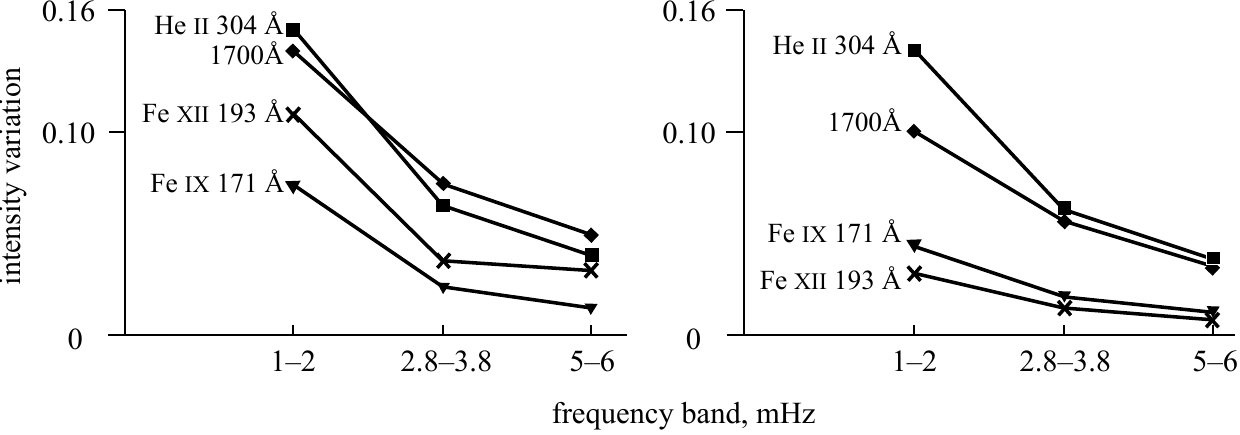}
}
\caption{Intensity variation at the specified frequency band for four UV lines 
(normalized to unity). Left panal -- facula No.\,1, right panel -- facula 
No.\,2}
\label{fig:variations-vs-freq}
\end{figure}

 We performed a similar analysis for a loop connecting Facula~1 and sunspot
NOAA\,11311 (Figure\,\ref{fig:sunspot-fac-171-2d}). Note that the oscillation spectra in the loop's footpoints
1 and 2 differ noticeably (Figure\,\ref{fig:fft-11311-4-points}, the upper row). The spectra of the elements
located fairly close to each other in the middle part of the loop show
differences as well (Figure\,\ref{fig:fft-11311-4-points}, lower row). Phase-speed measurements based
on the signals at points 3 and 4 in the middle part of the loop are also
ambiguous and yield values of 100\,--\,120\,km\,s$^{-1}$. One can conclude that oscillation
spectral-phase characteristics in the analyzed loop of Facula~1 considerably
differ from those of Facula\,2, which is caused by the NOAA\,11311 influence.
Figure\,\ref{fig:variations-vs-freq} represents the maximum intensity variation values for three
frequency bands at five heights. Both faculae show that the low frequencies
dominate at all heights, and the largest intensity variation is detected in
the transition region (the He\,\textsc{ii}\,304\,\AA\ line).



\section{Conclusions} 
      \label{S-Conclusions}
Low-frequency oscillations (1\,--\,2\,mHz) in sunspots are concentrated in penumbrae forming
annular areas, which expand with height. High-frequency oscillations
(5\,-–\,7\,mHz) are concentrated in fragments located in areas confined to the photosphere
umbra boundaries at all heights. Spatial distribution of low-frequency oscillations
in the corona above sunspots and faculae reproduces the coronal
fan structures well. This signifies that the upper parts of most coronal loops
conducting 10\,--\,15\,min oscillations are located within the Fe\,\,\textsc{ix} 171\,\AA\ 
line-formation layer, while three-minute and shorter-period oscillations possibly penetrate
to higher coronal levels by other loops. The observation-based dominant frequency \textit{vs.}
distance from barycenter relations may be used to determine inclination of magnetic tubes in
higher levels where it cannot be measured directly. The calculations show
that this angle is close to \textbf{40$\degree$} above the umbral borders in the transition region. 
Phase speeds measured in
the coronal loops' upper parts at the Fe\,\textsc{ix} 171\,\AA\ line formation height reach
100\,--\,150\,km\,s$^{-1}$ for sunspots and 50\,--\,100\,km\,s$^{-1}$ for faculae.
Intensity and LOS-velosity oscillation power spectra
differ significantly in the facular lower atmospheric layers: spectra of
intensity oscillations show the prevalence of the 0.7\,--\,1.2\,mHz frequency
oscillations; in those of LOS-velocity, the dominant oscillations are
five-minute oscillations. Such spectra are typical for both ground-based and space
observations. The absence of low-frequency peaks in LOS-velocity spectra may
signify that two oscillation types coexist in the region under study. Amplitude of intensity
oscillations is maximum in the transition region (He\,\textsc{ii} 304\,\AA) above
faculae.

\hyphenation{Pro-jects}
\small{\textbf{Acknowledgements}.
The study was performed with partial support of the Projects No.\,16.3.2, 16.3.3 
of ISTP SB RAS. We acknowledge E.~Korzhova for her help in preparing the English 
version of the article and the NASA/SDO science team for providing the data. We 
are grateful to an anonymous referee for the helpful remarks and suggestions. 


\bibliographystyle{spr-mp-sola}

\tracingmacros=2
\bibliography{kobanov}


\end{document}